\documentclass[aps,preprint,amsmath,amssymb]{revtex4}

\usepackage{graphicx}
\usepackage{color}

\def\g5{\gamma_{5}}
\def\ga{\gamma}
\def\ve{\varepsilon}

\def\be{\begin{eqnarray}}
\def\ed{\end{eqnarray}}
\def\t{\tilde}
\def\vectrl #1{\buildrel\leftrightarrow \over #1}
\def\partrl{\vectrl{\partial}}
\def\non{\nonumber}

\begin{document}
\title{\Large \bf  U-boson and the HyperCP exotic events }
\date{\today}

\author{ \bf  Chuan-Hung Chen$^{1,2}$\footnote{Email:
physchen@mail.ncku.edu.tw}, Chao-Qiang Geng$^{3}$\footnote{Email:
geng@phys.nthu.edu.tw}, and Chung-Wen Kao$^{4}$\footnote{Email:
cwkao@phys.cycu.edu.tw}
 }

\affiliation{ $^{1}$Department of Physics, National Cheng-Kung
University, Tainan 701, Taiwan \\
$^{2}$National Center for Theoretical Sciences, Hsinchu 300, Taiwan\\
$^{3}$Department of Physics, National Tsing-Hua University, Hsinchu
300, Taiwan  \\
$^{4}$Department of Physics, Chung Yuan Christian University,
Chung-Li 320, Taiwan }

\begin{abstract}
We show that the very light spin-1 gauge U-boson of the extra
$U(1)'$ gauge model in the framework of the supersymmetric  standard
model extension can be a good candidate of the new light particle
suggested by the HyperCP experiment. We  demonstrate that the flavor
changing neutral currents (FCNCs) for the HyperCP events in the
decay of $\Sigma^{+}\to p \mu^{+} \mu^{-}$ can be generated at both
tree and loop levels. In particular, we find that the loop induced
$s\to d U$ transition due to the tensor-type interaction with the
dimension-5 electric dipole operator plays a very important role on
the FCNCs. Our explanation of the HyperCP data with the spin-1
U-boson is different from that based on a light pseudoscalar Higgs
boson or sgoldstino in the literature. In particular, the U-boson
involves a rich phenomenology in particle physics as well as
cosmology.
\end{abstract}
\maketitle

An unexpected large branching ratio (BR) for $\Sigma^{+}\to p
\mu^{+} \mu^{-}$ has been measured to be $[8.6^{+6.6}_{-5.4}({\rm
stat})\pm 5.5({\rm syst})]\times 10^{-8}$ by the HyperCP experiment
at Fermilab \cite{HyperCP}. Due to the short-distance contributions
being negligible, this ``anomalous'' result could just reflect the
uncertain long-distance effects \cite{BSS,HTV_PRD72}. However, the
unforeseen result is actually based on three observed events, which
all appear at the narrow range of the dimuon mass distribution in
$\Sigma^{+} \to p \mu^{+} \mu^{-}$. The probability for the three
events arising from the form-factor decay spectrum in the standard
model (SM) is estimated to be around 0.8\% \cite{Kaplan}. Therefore,
a more accessible speculation is that a new neutral resonance,
denoted by $X^0$, with mass of $214.3\pm 0.5$ MeV has been produced
through the decay chain $\Sigma^+\to p X^0, X^0\to \mu^{+} \mu^{-}$
and the corresponding BR is ${\cal B}(\Sigma^+\to p X^0, X^0\to
\mu^{+} \mu^{-})=[3.1^{+2.4}_{-2.9}({\rm stat})\pm 1.5({\rm
syst})]\times 10^{-8}$ \cite{HyperCP}.

To be consistent with the HyperCP events, this new particle is
assumed to be light and weakly couple to the SM particles.
Subsequently, several possible candidates for this new particle have
already been suggested, such as a light pseudoscalar Higgs boson in
the next minimal standard supersymmetric model (NMSSM)
\cite{theory1} and a light sgoldstino in some supersymmetric models
\cite{sgoldstino}. In addition, a model-independence effective
interactions via scalar, pseudoscalar, vector and axial vector
currents have also been analyzed in Refs.
\cite{theory2,CG_PLB645,theory3}. In this study, besides giving a
specific model to realize the conjecture on axial vector currents
which have been studied by the model-independent approach,
 we will show that the tensor-type effective
interactions missed in the literature are also important for the
HyperCP data.

In the SM, the flavor changing neutral currents (FCNCs) are
generated by quantum loops. Consequently, it is believed that due to the
loop suppression and Glashow-Iliopoulos-Maiani (GIM) mechanism
\cite{GIM},  the associated FCNC processes are usually sensitive to
the new physics effects. It has been known that there exist many
extensions of the SM, such as supersymmetric \cite{SUSY}, left-right
symmetric  \cite{LRmodel} and flavor-changing $Z'$ \cite{Z0,Zprime}
models, which
all  involve new flavor structures and induce new FCNCs at loop
and/or tree levels. After surveying the models in the literature, we
find that the simplest extension of the SM, that could naturally
provide a light spin-1 boson and axial couplings to the SM fermions,
is the supersymmetrized $U(1)'$ gauge model.

It has been shown that a new neutral gauge boson associated with an
extra $U(1)'$ gauge symmetry is a necessity to be responsible for
the spontaneous supersymmetry breaking and the generation of large
sfermion masses \cite{Fayet1}. To distinguish from the normal
$Z'$-boson, here we use $U$-boson to represent the new gauge boson.
Since the U-boson is regarded as the spin-1 superpartner of the
massless spin-1/2 goldstino, its features include (a) axial and weak
couplings to the fermions and (b) a very light mass \cite{Fayet2}.
The special characters of the U-boson as well as its related
phenomenologies have been discussed extensively in
Refs.~\cite{Fayet3,Fayet4,Fayet5,Fayet6}. In addition, the studies
of directly detecting the U-boson at BESIII and identifying it as
$X^0$ can be found in Ref.~\cite{BES3}. As to other possible light
gauge boson physics, one could refer to Refs.~\cite{GK,Dob}. In this
paper, we are going to demonstrate that the spin-1 $U$-boson in the
supersymmetric $U(1)'$ model could be the candidate of the new light
particle $X^0$. In particular, we will show that the model provides
the mechanism of FCNCs in $\Sigma^{+}\to p\mu^{+} \mu^{-}$ for the
HyperCP events at both tree and loop levels. We emphasize that the
loop induced $s\to d U$ transition involves the electric dipole type
of tensor interactions, which has not been investigated yet in the
literature.

To make the model to be more concretely, we adopt the simplest
approach proposed in Ref.~\cite{Fayet2}, in which the theory is
based on the framework of supersymmetry (SUSY) with $SU(3)_c\times
SU(2)_{L}\times U(1)_{Y}\times U(1)'_X$ gauge symmetries. To get
correct symmetry breaking, the model involves two Higgs doublets
$H_{u, d}$ and one singlet $N$. Since the couplings between the
U-boson and SM fermions are very weak,  with the $\nu_{\mu} e$
scattering, the mass of the U-boson could
be less than a few hundreds MeV. In our model, since the couplings
of the U-boson to SM fermions are through axial vector currents, in
order to make the model be anomaly free, it is necessary to
introduce new heavy fermions with opposite $U(1)'$ charges to
ordinary fermions. Moreover, the $U(1)'$ charges of quarks would be
different for different generations like the nonuniversal
$Z^{\prime}$ model \cite{Z0,Zprime}. It should be noted that the
fermion $U(1)'$ charges for the models in the literature
\cite{Fayet1,Fayet2,Fayet3,Fayet4,Fayet5,Fayet6,BES3} are generation
blind and therefore, the U-boson in our discussion here should be
viewed as a variation of the conventional U-boson in the literature.
Nevertheless, the main features of the U-boson are the same.

To investigate the U-boson effects, we start by writing the
interactions of the U-boson and fermions as
 \be
 {\cal L}_{ f fU}&=& -  g_U \sum_{f} Q_{f}
 \left[ \bar f_R\ga_{\mu} f_R-\bar f_L \ga_\mu f_L
 \right]U^{\mu} \label{eq:effL}\,,
 \ed
where $g_{U}$ denotes the gauge coupling of $U(1)'$, $Q_f$ is the
 $U(1)'$ charge for the corresponding particle and $f_{L(R)}=P_{L(R)}f$ with
$P_{L(R)}=(1\mp \ga_5)/2$.
As an illustration, we may take $Q_2=-Q_1=q$ and $Q_3=0$, where 1, 2 and 3 are the family index.
 In this simple toy model, no more exotic fermions are needed as the $U(1)'$ related anomalies are
 cancelled among the first and second
families and the Yukawa couplings are ok, while Eq. (\ref{eq:effL}) is obtained. Of course, to have a realistic CKM mixings, a more
complex Higgs structure is needed.
 From Eq.~(\ref{eq:effL}), since gauge
charges are generation dependent,
FCNC
interactions can be generated by
 \be
g_{U}\bar D_i \ga_{\mu}\left( V^{D}_{R} \mathbf{Q} V^{D^\dagger}_{R}
P_R - V^{D}_{L} \mathbf{Q} V^{D^\dagger}_{L} P_L\right)_{ij} D_j
U^{\mu}\,, \label{eq:fcnc}
 \ed
where $D_i$ represents the physical eigenstate of the down quark,
${\rm diag}\mathbf{Q}_i=Q_i$ and $V^{D}_{L(R)}$ are the unitary
matrices for diagonalizing the down-type-quark Yukawa matrix
$Y^{d}$. We note that the flavor number of down quarks in the
anomaly-free model could be more than three. From the above result,
it is clear that although Eq.~(\ref{eq:effL}) contains only axial
couplings, due to the misalignment between $V^{D}_{L}$ and
$V^{D}_{R}$, in general the effective interactions of the U-boson
and quarks have not only axial vector currents but also vector ones.

It has been known that without finetuning, the CP violating phases
in models with SUSY bring a large effect on the neutron electric dipole
moment (NEDM). To solve this CP problem, the Yukawa and SUSY
soft breaking matrices could be considered as hermitian \cite{ABKL},
{\it i.e.}, $Y^{d^\dagger}=Y^{d}$ and $A^{d^\dagger}=A^{d}$,
respectively. Interestingly, the hermitian $Y^{d}$ could be
naturally realized in the extensions of the SM, such as left-right
symmetric models \cite{hermitian_y}. Moreover, with hermitian Yukawa
matrices, one can show that in the supergravity framework, $A^{d}$
is nearly hermitian even after including renormalization group
running effects \cite{ABKL}.
Despite the
origin of the hermiticity, if we adopt a hermitian matrix of
 $Y^{d}$, since
$M^{dia}_{D}= V^{D}_{L} Y^{d} V^{D^\dagger}_{R}$, immediately we get
$V^{D}_{L}=V^{D}_R\equiv V^{D}$. Accordingly, from Eq.~(\ref{eq:fcnc})
the FCNC for $s\to d U$ at tree
level is found to be
 \be
{\cal L}&=& -g_{U} V_{12}  \bar d \ga_{\mu}\ga_5 s U^{\mu}+h.c.
\label{eq:int_sd}
 \ed
where $V_{12}=(V^{D} \mathbf{Q} V^{D^\dagger})_{12}$. Clearly, we
successfully obtain the FCNCs at tree level in the specific model.
Meanwhile, we find that when Yukawa matrices have the property of
hermiticity, axially coupled effective interactions between the
U-boson and quarks in the physical states are returned as
Eq.~(\ref{eq:effL}). It should be an interesting problem to ask how
reliable the hermitian Yukawa matrices are. Moreover, it is worth
discussing about the mass matrices of quarks. It has been known that
the determination of flavor mixing matrices $V^{F}_{L, R}$ (F=D, U)
is governed by the detailed patterns of the mass matrices. In terms
of data, the CKM matrix, defined by $V^{U}_{L} V^{D^\dagger}_{L}$,
is approximately a unity matrix. Accordingly, the quark mass
matrices are very likely aligned and have the relationship of ${\cal
M}_{D}={\cal M}_{U} + \Delta(\lambda^2)$ with ${\cal
M}_{U(D)}=M_{U(D)}/m_{t(b)}$ \cite{qm1,qm2,qm3}, where $\lambda$ is
the Wolfenstein parameter. In Ref.~\cite{qm3}, it showed that the
Fritzsch quark mass matrices, given by \cite{Fritzsch,qm2} \be
M_{F}=R_{F} \bar M_{F} H_{F}\ { \rm with}\ \bar M_{F}=\left(
        \begin{array}{ccc}
          0 & A_{F} & 0 \\
          A_{F} & 0 & B_{F} \\
          0 & B_{F} & C_{F}  \\
        \end{array}
      \right) \label{eq:mass}
\ed
 where $R_{F}$ and $H_{F}$ are diagonal phase matrices, could
lead to reasonable structures for the mixing angles and CP violating
phase in the CKM matrix just in terms of the quark masses.
Interestingly, when $R_F=H^{\dagger}_F$,  the simple Fritzsch mass
matrices are hermitian. That is, although hermitian mass matrices of
quarks are a subset of general cases,
 the simple patterns
have brought us enough information for the flavor physics. Hence, in
our following analysis we will adopt the assumption of hermiticity
for the Yukawa matrices. We remark that the interaction in Eq.
(\ref{eq:int_sd}) can be easily extended to those with $b\to s$ and
$b\to d$ transitions~\cite{CG_PLB645}.

 From Eq.~(\ref{eq:int_sd}), we aware that FCNCs at tree
with the axial coupling for the $s\to d $ transition in the
literature could arise from the nonuniversal supersymmetric $U(1)'$
model. In addition, because the interacting form of
Eq.~(\ref{eq:int_sd}) is the same as that parametrized by the
model-independent approach. We can take the procedure discussed in
the Refs.~\cite{theory1,theory2,CG_PLB645,theory3} to constrain the
parameter $g_U V_{12}$. Consequently, if the events of the HyperCP
data are regarded as the production of the resonance, with the
narrow width approximation, the BR for $\Sigma^{+} \to p U, U\to
\mu^{+} \mu^{-}$ can be written as the product of ${\cal
B}(\Sigma\to p U)\times {\cal B}(U\to \mu^{+} \mu^{-})$. Since the
anomalous events show up only in the dimuon mode, it is plausible to
take ${\cal B}(U\to \mu^{+} \mu^{-})\approx 1$. We remark that in
general the $U$ particle would also couple to the electron and
neutrinos. In this case,  as the experimental constraints on Eq.
(\ref{eq:int_sd}) from $K_L\to \ell\bar{\ell}\ (\ell=e,\nu)$ are
much weaker than $K_L\to\mu^+\mu^-$, our numerical results in the
followings need to be simply rescaled.  Hence, using
Eq.~(\ref{eq:int_sd}) and the results of the Chiral Lagrangian for
the $\Sigma^{+} \to p U$ transition, by fitting the HyperCP data one
easily finds that $|g_UV_{12}|^2 \approx (4.4^{+3.4}_{-2.7}\pm
2.1)\times 10^{-20}$ \cite{theory2}.

To take the interaction in Eq. (\ref{eq:int_sd}) more seriously, we
should examine whether other experiments will give a more stringent
constraint on $g_{U}V_{12}$. It has been analyzed that in fact,
except $\Sigma^+ \to p \mu^{+} \mu^{-}$, the most serious bound is
from $K_{L}\to \mu^{+} \mu^{-}$ instead of the $K^0-\bar K^0$ mixing
\cite{theory2,CG_PLB645}.
 From the results in
Ref.~\cite{CG_PLB645} and by adopting ${\cal B}(K_L\to \mu^{+}
\mu^{-})< 10^{-10}$, one obtains $|g_{U} V_{12}|^2 \Gamma(U\to
\mu^{+} \mu^{-})< 2.8\times 10^{-30}$ with
 \be
\Gamma(U\to \mu^{+} \mu^{-})&=&\frac{|g_{U} Q_{\mu}|^2 m_{U}}{12\pi}
\left(1-\frac{4m_{\mu}^2}{m^2_{U}}\right)^{3/2}, \label{eq:rate}
 \ed
where $Q_{\mu}$ is the $U(1)'$ gauge charge of the muon.
Fortunately, the unknown parameter $g_{U} Q_{\mu}$ can be directly
constrained by the muon anomalous magnetic moment. Thus, the U-boson
mediated muon $g-2$ can be calculated to be \cite{Fayet6,g2}
 \be
 \Delta a_{\mu}=\frac{g^2_{\mu}}{4\pi^2}
\frac{m^2_{\mu}}{m^2_{U}} F_{U}\left(
\frac{m^2_{\mu}}{m^2_{U}}\right)\non
 \ed
with $g_{\mu}=g_{U} Q_{\mu}$ and
 \be
F_{U}(a)=\int^{1}_{0} dz { z(1-z)(4-z)+2az^3\over 1-z + a z^2}\,.
\non
 \ed
According to the current data, we know that the difference between
the experimental value and the SM prediction is $\Delta a_{\mu}=a^{\rm
exp}_{\mu}-a^{\rm SM}_{\mu}=(22\pm 10)\times 10^{-10}$ \cite{PDG06}.
Using $\Delta a_{\mu}< 10^{-9}$ and Eq.~(\ref{eq:rate}), the limit
on the partial decay rate is given by $\Gamma(U\to \mu^{+} \mu^{-})<
5.5\times 10^{-12}$. Combining with the bound of $K_L\to \mu^{+}
\mu^{-}$, we get
 \be
|g_{U} V_{12}|^2 < 5 \times 10^{-19}.
\label{eq:5}
 \ed
Obviously, $\Sigma^{+}\to p \mu^{+} \mu^{-}$ itself provides
stronger constraint on the flavor changing parameter when  $BR(U\to
\mu^{+} \mu^{-})\approx 1$ is assumed.

So far, we have just paid attention to the effects of
Eq.~(\ref{eq:int_sd}), arising from the FCNCs at tree level via
axial-vector current interactions. Next, we will introduce another
interesting tensor type dipole operator which has been missed in the
literature and can have significant contributions to the HyperCP
data. To introduce the new type interaction for $s\to d \mu^+ \mu^-$
in a model-independent way and avoid the strong constraint from
$K\to \pi \mu^{+} \mu^{-}$, we first parametrize the new interaction
to be a dimension-5 dipole operator, given by
 \be
{\cal L}_{T^5}=-\frac{g_{T^5}}{2\Lambda_{N}} \bar d\; i \sigma^{\mu
\nu} \g5 s
  X^{0}_{\mu\nu}+h.c.,
  \label{eq:dipole}
 \ed
where $\Lambda_{N}$ denotes the energy scale of new physics,
$g_{T^5}$ is a dimensionless parameter and $X^{0}_{\mu
\nu}=\partial_{\mu} X^{0}_{\nu} -
\partial_{\nu} X^{0}_{\mu}$.
We note that the interaction in Eq. (\ref{eq:dipole}) gives no
contribution to $K\to \pi\mu^{+}\mu^{-}$ due to the parity
conservation in strong interaction and moreover, it does not
contribute to $K_{L}\to \mu^{+} \mu^{-}$ either, unlike that with
the axial-vector type interaction.
 In terms of $\langle p | \bar d \sigma_{\mu\nu}
\g5 s| \Sigma \rangle=c_{\sigma} \bar p \sigma_{\mu\nu}\g5 \Sigma$
and $c_{\sigma}=-1/3$ \cite{DGH,HTV_PRD72}, the transition matrix
element for $\Sigma^{+} \to p X^0$ is written as
 \be
  M(\Sigma^{+}\to p X^0)&=& \frac{g_{T^{5}}}{\Lambda_{N}}  c_{\sigma}\bar{p}\, i\sigma_{\mu \nu}
  q^{\nu}  \g5 \Sigma\, \ve^{\mu*}_{X^0} \non
 \ed
 and the transition amplitude square is obtained by
 \be
  |M(\Sigma^{+}\to p X_A)|^{2}&=& 4 m_{\Sigma} \left( g_{T^{5}}
  c_{\sigma}\right)^2
   \left[4E_{X} p_{p}\cdot p_{X}-m^2_{X} E_{p}+3 m_{p} m^2_{X} \right]\,.\non
 \ed
 Consequently, the BR for the decay chain is found to be
 \be
  &&{\cal B}(\Sigma^{+}\to p X^0, X^0\to \mu^{+} \mu^{-})\non\\
  && =2.2\times
  10^{11}g_{T^{5}}^2  \left(\frac{\rm GeV}{\Lambda_{N}}\right)^2 {\cal B}(X^0\to \mu^{+}
  \mu^{-})\,. \non
 \ed
If the anomalous HyperCP events  are dictated by the electric dipole
operator in Eq.~(\ref{eq:dipole}), we find
  \be
   \left(\frac{g_{T^{5}}}{\Lambda_N}\right)^2 {\cal B}(X^0\to \mu^{+}
   \mu^{-})&=&\left(1.41^{+1.09}_{-0.86} \pm 0.68 \right) 10^{-19} GeV^{-2}\,. \label{eq:vgt5}
  \ed
With ${\cal B}(X^0\to \mu^{+} \mu^{-})\approx 1$,
Eq.~(\ref{eq:vgt5}) could be regarded as the bound on
$g^2_{T^5}/\Lambda^2_N$ model-independently.

\begin{figure}[htbp]
\includegraphics*[width=3.5in]{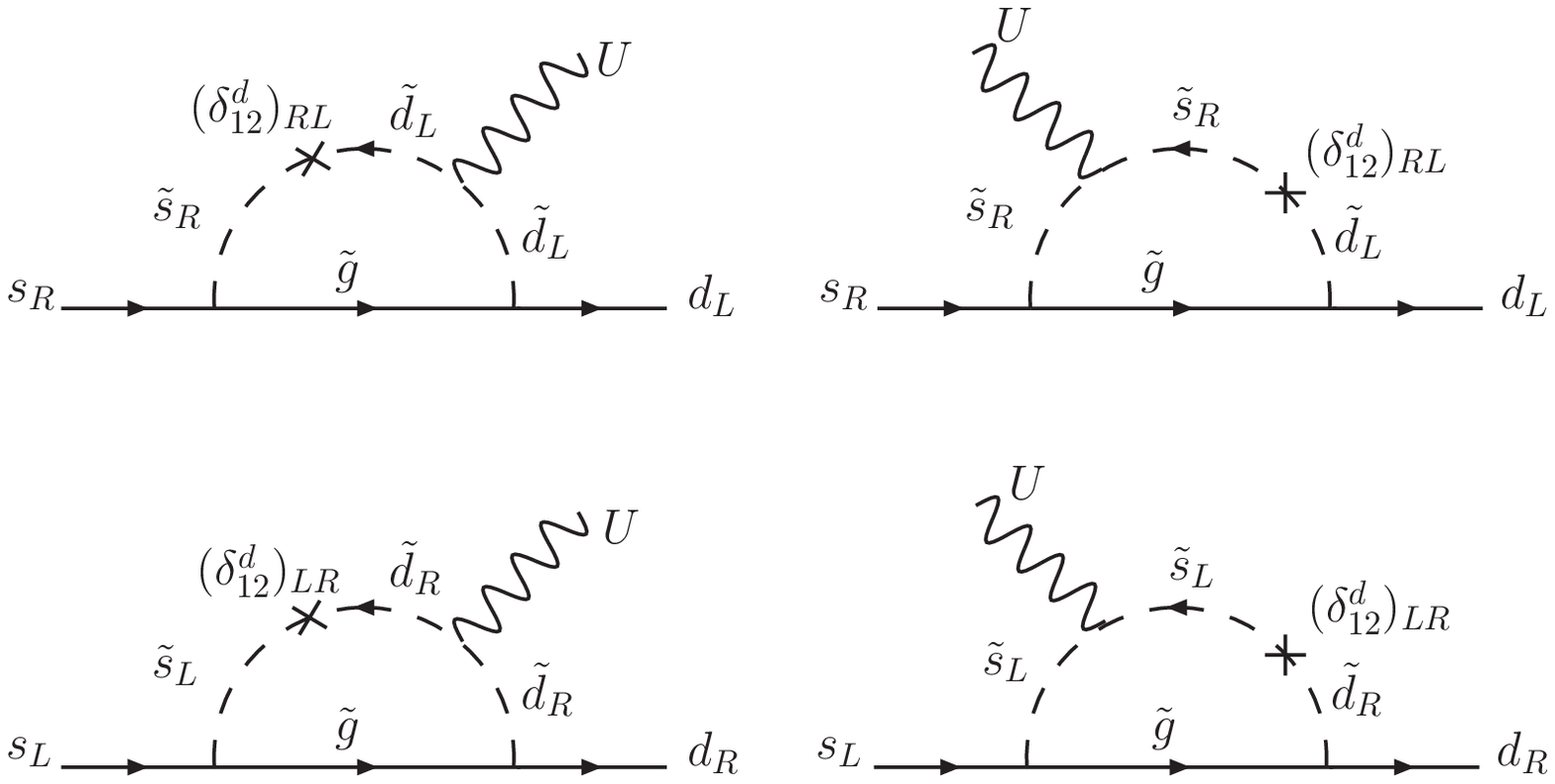}
\caption{U-penguin diagrams
for $s\to d U$. }
 \label{fig:U-pen}
\end{figure}

After analyzing the importance of the dimension-5 dipole operator,
the question is how to construct a physical model to satisfy the
condition in Eq.~(\ref{eq:vgt5}).  In what follows, we are going to
demonstrate that the new type operator in Eq.~(\ref{eq:dipole})
could be realized in the U-boson model. Since  the R-parity in our
consideration is conserved, to examine the U-boson effects in the
SUSY framework of models, we also need to know the couplings of the
U-boson and squarks. In terms of SUSY, from Eq.~(\ref{eq:effL}) the
interactions of the U-boson to squarks are found to be
 \be
 {\cal L}_{\t f \t fU}&=& - i g_U \sum_{f} Q_{f}
 \left[ \t f_R^*\partrl_\mu \t f_R-\t f_L^*\partrl_\mu \t f_L
 \right]U^{\mu} \label{eq:effSU} \,. \non
 \ed
Clearly, to get
the $s\to d U$ transition, we need to
calculate the U-penguin diagrams illustrated in
Fig.~\ref{fig:U-pen}.
In the figure, the mass insertion parameters
$(\delta^{d}_{ij})_{LR}$ are defined by
 \be
(\delta^{d}_{ij})_{LR}&=& \frac{1}{m^2_{\t
q}}\left(A^{d^\dagger}_{ij} v_{d} - Y^{d}_{ij} \mu v_u \right)\,,
 \ed
where $m_{\t q}$ is the average squark mass, $v_{d(u)}$ is the
vacuum expectation value (VEV) of $H_{d(u)}$  and $\mu$ is the
mixing parameter of $H_{d}$ and $H_{u}$.

As usual, we employ the mass insertion approximation to estimate
the contributions in Fig.~\ref{fig:U-pen} and the result is obtained to be
proportional to
 \be
 f_{sd} \bar d i\sigma_{\mu\nu} \left[
 \left(\delta^{d}_{12}\right)_{RL} P_R
 - \left(\delta^{d}_{12}\right)_{LR} P_L \right]sU^{\mu\nu}
 \label{eq:dim5}
  \ed
with $U^{\mu \nu}=\partial ^{\mu } U^{\nu} - \partial^{\nu} U^{\mu}$
and $f_{sd}=g_{U}(Q_{s}-Q_{d})$. Here, the
super-Cabibbo-Kobayashi-Maskawa (SCKM) basis has been adopted,
{\em i.e.}, the squark fields have been transformed into the states that
 $Y^{d}$ is diagonalized. In addition, the interactions for the
gluino-quark-squark are taken to be
 \be
 {\cal L}&=&-\sqrt{2} g_s \left[\bar q P_R \t g^a T^a \t q_L
  - \bar q P_L \t g^a T^a \t q_R \right] +h.c.\,, \non
 \ed
where $g_s$ is the strong coupling constant and $T^q$ denotes the
Gell-Mann matrices.
Although the U-boson is axially coupled to
quarks in the interacting eigenstates and its couplings to the
left-handed and right-handed squarks are the same (opposite)
in magnitude
(sign), due to the misalignment between quarks and squarks in the
physical states, the transition amplitudes for $s\to d U$ induced by
the U-penguin diagrams involve not only electric dipole  but also
magnetic dipole operators. In the early analysis, we have shown that a
hermitian $Y^{d}$ could naturally lead to pure axial-vector
couplings. Moreover, the hermiticity of $Y^{d}$ also makes the
$A^{d}$ be nearly hermitian. Hence, by utilizing the hermiticities of
$Y^{d}$ and $A^{d}$, the mass insertion parameters could be
simplified to be
 \be
   (\delta^{d}_{12})_{RL} \approx (\delta^{d}_{12})_{LR}\,.
   \label{eq:lreq}
 \ed
It is worth mentioning that with the same requirement, one can also
find that CP asymmetries in $\Lambda\to p \pi $ decays could
naturally be as large as $O(10^{-4})$ where the SM prediction is
$O(10^{-5})$ \cite{Chen_PLB521}.
With the property in Eq.~(\ref{eq:lreq}), the exact transition of
$s\to d U$ in Eq.~(\ref{eq:dim5}) can be written in terms of the
dimension-5 operator as
  \be
     {\cal L}_{dim5}=-\frac{g_5}{2m_{\t g}} \bar d i\sigma_{\mu \nu} \gamma_5 s U^{\mu
     \nu} +h.c. \label{eq:U-dipole}
  \ed
where
 \be
     g_5&=& \frac{g^2_{s} C_{F}}{16\pi^2}
     f_{sd} x {(\delta^{d}_{12})_{RL}}
     \label{eq:g5}
 \ed
with $C_F=4/3$, $x=m^2_{\t g}/m^2_{\t q}$ and $m_{\t g}$ being the
 gluino mass. By comparing to Eq.~(\ref{eq:dipole}),  here the new
physics scale can be identified to be the SUSY breaking related scale, {\it i.e.}
$\Lambda_{N}=m_{\t g}$.

\begin{figure}[hpbt]
\includegraphics*[width=3. in]{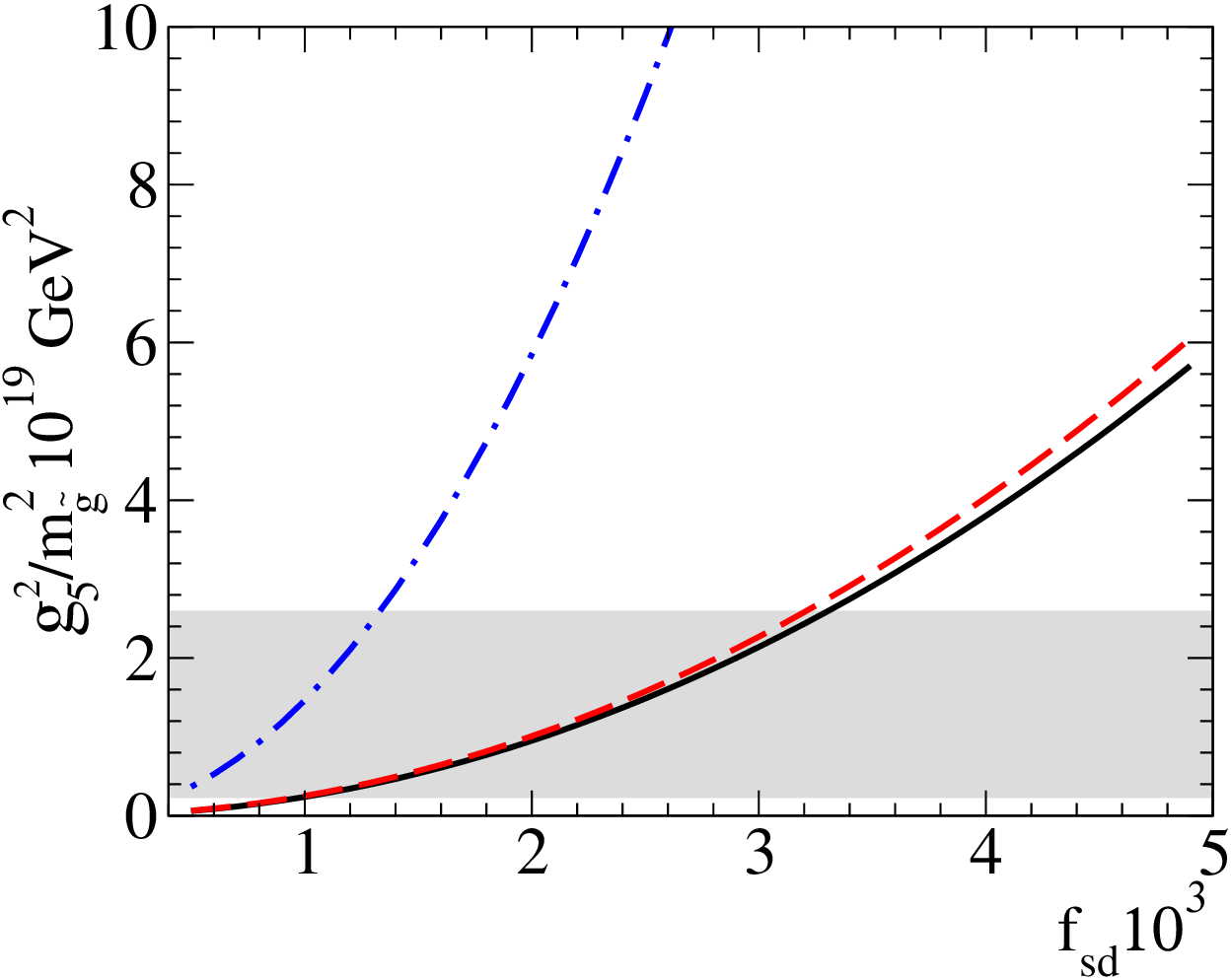}
\caption{$g^2_5/m^2_{\t g}$
as a
function of $f_{sd}$,
where the solid, dashed and
dot-dashed lines denote $x=0.3$, $1.0$ and $4.0$, respectively, and
the band is the HyperCP data with $1\sigma$ errors. }
 \label{fig:vg52}
\end{figure}

We now examine the naturalness for $g_5$ to satisfy the limit of the
HyperCP data given by Eq.~(\ref{eq:vgt5}).
 From Eq.~(\ref{eq:g5}), we see that the
unknown parameters are the average gluino and squark masses,
$f_{sd}$ and $(\delta^{d}_{12})_{RL}$. It is known that
$(\delta^{d}_{12})_{RL}$ associated with a specific value of $x$
could be constrained by the $K^0-\bar K^0$ mixing. According to the
results in Ref.~\cite{GGMS}, we present the constraints in
Table~\ref{tab:delta}.
%
\begin{table}[hptb]
\caption{ The constraints on $(\delta^{d}_{12})_{RL}$ by $\Delta
m_{K}$
with $m_{\t q}=500$ GeV and
$x=m^2_{\t g}/m^2_{\t q}$ being $0.3$, $1.0$ and $4.0$, respectively
\cite{GGMS}. }\label{tab:delta}
\begin{ruledtabular}
\begin{tabular}{cccc}
$x$  & $0.3$ & $1.0$ & $4.0$
 \\ \hline
$(\delta^{d}_{12})_{RL}$ & $7.9\times 10^{-3}$ & $4.4\times 10^{-3}$
&
$5.3\times 10^{-3}$ \\
\end{tabular}
\end{ruledtabular}
\end{table}
%
 From the decays of
$\eta^{\prime} \to UU$ and $\eta \to UU$,  we can get the direct
constraints for $f_s=g_U Q_s$ and $f_d=g_U Q_d$ to be $ < 5 \times
10^{-2}$ and $ 3.18 \times 10^{-2}$, respectively \cite{Fayet5},
which lead to $f_{sd}=f_{s}-f_{d}< 3 \times 10^{-2}$. Since the
direct constraints on the parameters are looser, we take $f_{sd}$ as
a free parameter to  fit the HyperCP data. However, it is clear that
with a small value of $g_{5}$ as the one in Eq. (\ref{eq:5}), the
effect of Eq. (\ref{eq:U-dipole}) will disappear.
With
$m_{\t q}=500$ GeV, the value of $g^2_5$ as a function of $f_{sd}$
is presented in Fig.~\ref{fig:vg52}.
%
 From the results, we see clearly that the dimension-5 operator induced by the U-boson in supersymmetric models
also provides a plenty of allowed space (the band in
Fig.~\ref{fig:vg52}) for the HyperCP data. Explicitly, with
$f_{sd}\sim 2.5\times 10^{-3}$ and $x\sim 1$, we obtain
$g^2_5/m^2_{\t g}\sim 10^{-19}$ GeV$^{-2}$ which is within the
model-independent constraint shown in Eq.~(\ref{eq:vgt5}).

In summary, we have studied the  scenario of the very light gauge
U-boson which weakly couples to fermions in the framework of SUSY
with one extra $U(1)'$ gauge symmetry \cite{Fayet2}. We have shown
that the spin-1 U-boson can be a good candidate of the new light
particle suggested by the HyperCP experiment. In the model, we have
found that the FCNCs for the HyperCP events in the decay of
$\Sigma^{+}\to p \mu^{+} \mu^{-}$ can be generated at both tree and
loop levels. In particular, we have pointed out that the loop
induced $s\to d U$ transition, involving the tensor-type interaction
with the dimension-5 electric dipole operator, plays a very
important role on the HyperCP data. This interaction has not been
investigated previously in the literature as it gives no
contributions to $K_{L}\to \mu^{+}\mu^{-}$ and $K\to
\pi\mu^{+}\mu^{-}$. Finally, we remark that the contributions from
Eqs. (\ref{eq:int_sd}) and (\ref{eq:U-dipole}) to $\Sigma^{+}\to p
e^{+} e^{-}$ are negligible in comparison with that in the SM
\cite{HTV_PRD72}. The study of the tensor interaction has an impact
on the decay of $K_{L}\to\gamma\mu^{+}\mu^{-}$ and similar
discussions can be also generalized to $B$ and  $\tau$ decays
\cite{CG_PLB645}. Our explanation of the HyperCP events with the
spin-1 U-boson is clearly different from that based on a light
spin-0 pseudoscalar Higgs boson in Ref. \cite{theory1} or a light
sgoldstino in Ref. \cite{sgoldstino}. In particular, we emphasize
that the U-boson can involve a rich phenomenology in particle
physics as well as cosmology
\cite{Fayet1,Fayet2,Fayet3,Fayet4,Fayet5,Fayet6,BES3,GK,Dob}.\\

\begin{acknowledgments}

This work is supported in part by the National Science Council of
R.O.C. under Grant \#s:NSC-95-2112-M-006-013-MY2,
NSC-95-2112-M-007-059-MY3 and NSC-96-2112-M-033-003-MY3.
 \end{acknowledgments}

\end{document}